\documentclass[preprint,aps,prd,showpacs,superscriptaddress,nofootinbib,tightenlines]{revtex4}

\usepackage{amsfonts}
\usepackage{multirow}
\usepackage{mathrsfs}
\usepackage{graphicx}
\usepackage{amsmath}
\usepackage{amssymb}
\usepackage{bm}
\usepackage{bbm}
\usepackage{color}

\def\OMIT#1{}

\newcommand{\nn}{\nonumber}

\newcommand{\beq}{\begin{equation}}
\newcommand{\eeq}{\end{equation}}
\newcommand{\bqa}{\begin{eqnarray}}
\newcommand{\eqa}{\end{eqnarray}}

\begin{document}
\title{\mbox{}\\[10pt]
Fragmentation function of gluon into spin-singlet $\bm{P}$-wave quarkonium
}

\author{Feng Feng\footnote{F.Feng@outlook.com}}
\affiliation{China University of Mining and Technology, Beijing 100083, China\vspace{0.2cm}}
\affiliation{Institute of High Energy Physics, Chinese Academy of
Sciences, Beijing 100049, China\vspace{0.2cm}}

\author{Saadi Ishaq\footnote{saadi@ihep.ac.cn}}
\affiliation{Institute of High Energy Physics, Chinese Academy of
Sciences, Beijing 100049, China\vspace{0.2cm}}
\affiliation{School of Physics, University of Chinese Academy of Sciences,
Beijing 100049, China\vspace{0.2cm}}

\author{Yu Jia\footnote{jiay@ihep.ac.cn}}
\affiliation{Institute of High Energy Physics, Chinese Academy of
Sciences, Beijing 100049, China\vspace{0.2cm}}
\affiliation{School of Physics, University of Chinese Academy of Sciences,
Beijing 100049, China\vspace{0.2cm}}

\author{Jia-Yue Zhang\footnote{zhangjiayue14@ucas.ac.cn}}
\affiliation{School of Physics, University of Chinese Academy of Sciences,
Beijing 100049, China\vspace{0.2cm}}
\affiliation{Institute of High Energy Physics, Chinese Academy of
Sciences, Beijing 100049, China\vspace{0.2cm}}

\date{\today}
\begin{abstract}
Following the operator definition of the fragmentation function developed by Collins and Soper,
we compute the gluon-to-$h_c$ fragmentation function at the lowest order in the velocity expansion in
NRQCD factorization approach.
Utilizing some modern technique developed in the area of multi-loop calculation,
we are able to analytically deduce the infrared-finite color-singlet short-distance
coefficient associated with the fragmentation function.
The fragmentation probability for gluon into $h_c$ is estimated to be order $10^{-6}$.
\end{abstract}



\maketitle

Like parton distribution functions (PDFs), fragmentation functions (FFs) constitute one of
the fundamental probes to uncover the nonperturbative partonic structure related to a hadron.
According to the QCD factorization theorem~\cite{Collins:1989gx},
in a high-energy collision experiment with two colliding beams composed of hadrons of type $A$ and $B$,
the inclusive production rate of the identified hadron $H$ at very large transverse momentum,
is dominated by the fragmentation mechanism:
\beq
d\sigma[A+B\to H(P_\perp)+X]  = \sum_i d{\hat\sigma}[A+B\to i(P_\perp/z)+X] \otimes
D_{i \to H}(z,\mu)+{\mathcal O}(1/P_\perp^2),
\label{QCD:factorization:theorem}
\eeq
where $d\hat{\sigma}$ denotes the partonic cross section, the fragmentation function $D_{i \to H}(z)$
characterizes the probability for the parton $i$ to materialize into a complicated multi-hadron state that contains the hadron $H$
carrying the fractional light-cone momentum $z$ with respect to the parent parton.
The sum in (\ref{QCD:factorization:theorem}) is extended over all parton specifies ($i=q,\bar{q},g$).
Similar to the PDFs, the FFs are nonperturbative, but, universal objects,
whose scale dependence is governed by the Dokshitzer-Gribov-Lipatov-Altarelli-Parisi (DGLAP) equation.
Specifically speaking, the scale dependence of the gluon fragmentation function is controlled by
\beq
{d \over d \ln \mu^2} D_{g \to H}(z,\mu) =  \sum_{i} \int_z^1 {d\xi\over \xi}
P_{i g} (\xi, \alpha_s(\mu)) D_{i \to H}\left({z\over \xi},\mu\right),
\label{DGLAP:evolution}
\eeq
where $P_{ig}(\xi)$ designate the splitting kernel,
and $\mu$ is also frequently referred to as the QCD factorization scale,
since it also enters in the partonic cross section in (\ref{QCD:factorization:theorem}).
Once this FF  can be deduced at some initial scale $\mu_0$ by some means,
one can then determine its form at any other scale $\mu$ by solving the evolution equation
(\ref{DGLAP:evolution}).

In contrast to the fragmentation functions for light hadrons, the functions for a parton to fragmentate
into a heavy quarkonium, a nonrelativistic bound state composed of a
heavy quark and heavy antiquark, need not be viewed as genuinely
nonperturbative objects.
In fact, owing to the weak QCD coupling at the length scale $\sim 1/m$ ($m$ represents the heavy quark mass)
as well as the nonrelativistic nature of quarkonium,
the nonrelativistic QCD (NRQCD) factorization~\cite{Bodwin:1994jh} can be invoked to
refactorize the quarkonium FFs as
the sum of products of short-distance coefficients (SDCs) and long-distance yet universal
NRQCD matrix elements~\cite{Braaten:1993mp,Braaten:1993rw}.
To some extent, the profiles of quarkonium FFs are largely determined by perturbative QCD,
which renders the NRQCD approach a particularly predictive theoretical framework.
Recently, armed with various fragmentation functions computed in NRQCD approach,
a phenomenological analysis based on (\ref{QCD:factorization:theorem})
is conducted to confront with copious large-$P_\perp$ $J/\psi$, $\chi_{cJ}$ and $\psi^\prime$ data accumulated
at \textsf{LHC} experiments~\cite{Bodwin:2014gia,Bodwin:2015iua}.

Since the original computation of the quark and gluon fragmentation into $S$-wave quarkonium
using NRQCD approach by Braaten and collaborators~\cite{Braaten:1993mp,Braaten:1993rw},
numerous fragmentation functions for quark/gluon into various quarkonium states,
have been calculated in NRQCD approach during the past two decades~\cite{Braaten:1993jn,Ma:1994zt,Braaten:1994kd,Cho:1994qp,Ma:1995ci,Ma:1995vi,Braaten:1995cj,Cheung:1995ir,Qiao:1997wb,Braaten:2000pc,
Bodwin:2003wh,Sang:2009zz,Bodwin:2012xc,Artoisenet:2014lpa,Bodwin:2014bia,Gao:2016ihc,Sepahvand:2017gup,Zhang:2017xoj}.
For a recent compilation of the SDCs associated with various FFs,
we refer the interested readers to Ref.~\cite{Ma:2013yla}.

The goal of this work is to compute the FF of gluon into the spin-singlet $P$-wave quarkonium, exemplified by
the $h_{c,b}$ states. To date, these $P$-wave quarkonium states have only been
observed in the $e^+e^-$ collision experiments
via hadronic transitions from higher vector quarkonium states~\cite{Rosner:2005ry,Adachi:2011ji}.
It is conceivable that they will be established at LHC experiments in the future,
owing to enormous partonic luminosity there.
For this purpose, it is desirable if one can make accurate predictions for the fragmentation functions for
the $h_{c,b}$ states.

Due to the odd $C$ parity of $h_c$, the gluon fragmentation remnants must involve two additional gluons
in the color-singlet channel, considerably more complicated than gluon-to-$\chi_{cJ}$ FF considered in \cite{Braaten:1994kd,Ma:1995ci}, which only involves one additional gluon in the fragmentation products.
Although the Feynman diagrams are topologically identical to those for gluon-to-$J/\psi$ FF~\cite{Braaten:1995cj},
the actual calculation is much more challenging,
due to the occurrence of the IR divergence in the former case,
whereas the latter is free from IR singularity, at least up to the relative order $v^2$~\cite{Bodwin:2003wh}.

The gluon-to-$h_c$ fragmentation function was originally calculated by Hao, Zuo and Qiao in 2009~\cite{Hao:2009fa}.
Unfortunately, the authors of \cite{Hao:2009fa} seem not to employ a gauge-invariant regulator to
regularize the encountered IR singularity in the course of their calculation.
Also, their final results are expressed in terms of a two-fold integral with rather complicated integrand.

Very recently, there appears a notable technical progress in evaluating fragmentation functions, which borrows
some clever trick from higher-order calculation involving multi-body phase space integration~\cite{Zhang:2017xoj}.
The authors of \cite{Zhang:2017xoj} are able to compute the gluon-to-$J/\psi$ FF in a closed form,
which was unimaginable from the angle of the conventional method~\cite{Braaten:1995cj,Bodwin:2003wh,Bodwin:2012xc}.

Stimulated by the advance made in \cite{Zhang:2017xoj}, we feel that it is the time
to revisit the gluon-to-$h_c$ fragmentation function.
We will start from the gauge-invariant operator definition for this FF~\cite{Collins:1981uw},
and employ dimensional regularization (DR) as the gauge-invariant IR regulator.
Ultimately, we will also be able to achieve the analytical expression for this FF.

We choose to evaluate this gluon fragmentation function in a frame such that the $h_c$
has vanishing transverse momentum.
It is customary to adopt the light-cone coordinates in calculating FF.
Any four-vector $A^\mu=(A^0,A^1,A^2,A^3)$ can be recast in a light-cone format $A^\mu=(A^+,A^-, {\bf A}_\perp)$,
with $A^\pm \equiv {1\over \sqrt{2}}(A^0\pm A^3)$ and ${\bf A}_\perp \equiv (A^1,A^2)$.
The scalar product of two four-vectors $A$ and $B$ thus becomes
$A\cdot B=A^+B^- + A^-B^+ - {\bf A}_\perp \cdot {\bf B}_\perp$.
Specifically speaking, the four-momentum of the $h_c$ meson can be written as
$P^\mu = \big(P^+, P^- \equiv M_{h_c}^2/(2P^+),{\bf 0}_\perp \big)$, where $M_{h_c}\approx 2 m_c$ signifies the mass of
the $h_c$ meson.

A gauge-invariant operator definition for the fragmentation functions was
formulated by Collins and Soper in 1981~\cite{Collins:1981uw}.
This definition was first utilized by Ma to compute the quarkonium FFs in NRQCD~\cite{Ma:1994zt}.
For the intended $g$-to-$h_c$ fragmentation function,
the operator definition is given by~\cite{Collins:1981uw} (also see \cite{Bodwin:2003wh,Bodwin:2012xc}:
\bqa \label{CS:def:Fragmentation:Function}
& & D_{g \to h_c}(z,\mu) =
\frac{-g_{\mu \nu}z^{D-3} }{ 2\pi k^+ (N_c^2-1)(D-2) }
\int_{-\infty}^{+\infty} dx^- e^{-i k^+ x^-}
\\
&& \times
\langle 0 | G^{+\mu}_c(0)
\Phi^\dagger(0,0,{\bf 0}_\perp)_{cb}  \sum_{X} |h_c(P,\lambda)+X\rangle \langle h_c(P,\lambda)+X|
\Phi(0,x^-,{\bf 0}_\perp)_{ba} G^{+\nu}_a(0,x^-,{\bf 0}_\perp) \vert 0 \rangle,
\nn
\eqa
where $z$ denotes the fraction of the $+$-momentum carried by $h_c$ with respect to the gluon,
$D=4-2\varepsilon$ signifies the space-time dimensions, $N_c=3$ is the number of colors.
$G_{\mu\nu}$ is the matrix-valued gluon field-strength tensor in the adjoint representation of $SU(N_c)$,
$k^+ = P^+/z$ is the $+$-component momentum of injected by the gluon field strength operator.
$\mu$ is the renormalization scale for this composite operator.
The insertion of the intermediate states implies
that in the asymptotic future, one only needs project out those out-states that contain a
$h_c$ meson carrying definite momentum $P^\mu$ and polarization $\lambda$, plus any unobserved hadrons,
which are collectively denoted by the symbol $X$.
Note the summation on $X$ also implicitly indicates that
the three polarizations of $h_c$ are summed over.

The gauge link (eikonal factor) $\Phi(0,x^-,{\bf 0}_\perp)$ in (\ref{CS:def:Fragmentation:Function}) is a
path-ordered exponential of the gluon field, whose role is to ensure the gauge invariance of the FF:
\beq
\Phi(0,x^-,{\bf 0}_\perp)_{ba} = \texttt{P} \exp
\left[ i g_s \int_{x^-}^\infty d y^- n\cdot A(0^+,y^-,{\bf 0}_\perp) \right]_{ba},
\label{Gauge:Link:Definition}
\eeq
where $\texttt{P}$ implies the path-ordering, $g_s$ is the QCD coupling constant, and
$A^\mu$ designates the matrix-valued gluon field in the adjoint representation.
$n^\mu =(0,1,{\bf 0}_\perp)$ is a reference null 4-vector.
Actually, due to the $C$-odd property of the $h_c$ state, to our concerned perturbative order,
we do not need consider the complication for gluons to attach to the eikonal line.

According to the NRQCD factorization,
the gluon fragmentation function for $h_c$ can be expressed as~\cite{Hao:2009fa},
\beq
D_{g \to h_c}(z)  =  {d_1(z,\mu_\Lambda)\over m^5} \langle 0| {\cal O}_1^{h_c}(^1P_1)|0\rangle +
{d_8(z)\over m^3} \langle 0\vert {\cal O}_8^{h_c}(^1S_0)(\mu_\Lambda) \vert 0\rangle + \cdots,
\label{FF:NRQCD:factorization}
\eeq
where both color-singlet and color-octet channels contribute at lowest order in $v$.
The corresponding $h_c$ production operators in NRQCD are defined as~\cite{Bodwin:1994jh}~\footnote{
It was pointed out by Nayak, Qiu and Sterman~\cite{Nayak:2005rw,Nayak:2005rt} in 2005 that the original definition
of the NRQCD color-octet production operator is not gauge invariant, and the correct definition necessitates
the inclusion of eikonal lines that run from the quark/antiquark field to infinity.
To the perturbative order we are concerned, this nuisance can be safely neglected so we
stick to the conventional definition.}
\begin{subequations}
\label{hc:NRQCD:production:operators}
\bqa
\mathcal{O}_{1}^{h_c}(^1P_1)  &=&
\chi^\dagger \left(-\frac{i}{2}\tensor{\bf{D}}\right) \psi  \sum_{X} |h_c+X\rangle  \cdot \langle h_c+X|
\psi^\dagger \left(-\frac{i}{2}\tensor{\bf{D}}\right) \chi,
\\
\mathcal{O}_{8}^{h_c}(^1S_0) &=&
\chi^\dagger T^a\psi \sum_{X} |h_c+X\rangle \langle h_c + X|\; \psi^\dagger T^a\chi.
\eqa
\end{subequations}
The $\mu_\Lambda$ in (\ref{FF:NRQCD:factorization}) refers to the NRQCD factorization scale~\footnote{The scale $\mu_\Lambda$ here should not be confused with the QCD factorization scale
scale $\mu$ introduced in (\ref{CS:def:Fragmentation:Function}), which enters the DGLAP equation.}, which lies in the range
$m v \leq \mu_\Lambda \leq m$.
These two production operators are linked by the following renormalization group equation in NRQCD~\cite{Bodwin:1994jh}:
\begin{equation}
	\frac{d}{d\ln\mu^2_\Lambda}\langle {\cal O}_8^{h_c}(^1S_0)(\mu_\Lambda)\rangle = \frac{2 \alpha_s C_F}{3 \pi N_c m^2} \langle {\cal O}_1^{h_c}(^1P_1)\rangle,
\end{equation}
where $C_F=(N_c^2-1)/2N_c$.

\begin{figure}[tb]
\centering
\includegraphics[width=0.8\textwidth]{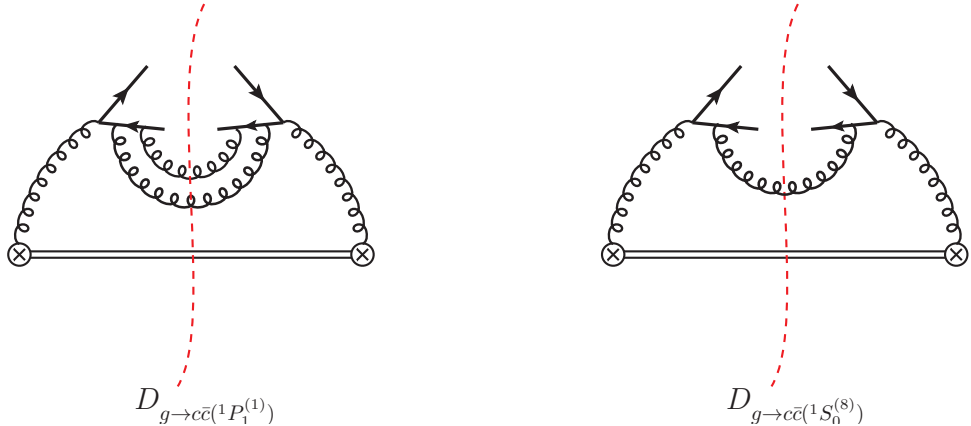}
  \caption{Representative Feynman diagrams for the gluon fragmentation function $D_{g\to h_c}(z)$.
  The double-line signifies the eikonal line, and the vertical dashed curve implies the Cutkosky cut.
  The cap represents the insertion of the operator $G_a^{+\mu}$, whose Feynman rule reads
  $+i \left(g^{\mu\alpha}-{Q^\mu n^\alpha\over Q\cdot n}\right)\delta_{ab}$ (left to the cut),
  where $Q$ represents the momentum of the gluon flowing out of the cap.
\label{Feynman:diagram}}
\end{figure}

We will proceed to compute the two SDCs using the standard perturbative matching technique, by replace the 
physical $h_c$ state by the free $c\bar{c}({}^1P_1^{(1)})$ and $c\bar{c}({}^1S_0^{(8)})$ states, respectively, 
in (\ref{FF:NRQCD:factorization}). The representative Feynman diagrams
are shown in Fig.~\ref{Feynman:diagram}. Since (\ref{CS:def:Fragmentation:Function}) 
is manifestly gauge invariant, for simplicity we adopt the Feynman gauge. Dimensonal regularization will be used
to regularize both UV and IR divergences.
It is convenient to employ the well-known covariant projector technique to expedite the calculation~\cite{Petrelli:1997ge}.
Since one does not bother to consider the situation where gluons attach to the eikonal line, it becomes rather
straightforward to generate the squared quark-level amplitude.
We employ the package \textsf{QGraf}~\cite{Nogueira:1991ex} to generate the corresponding Feynman diagrams and amplitudes,
then use the packages \textsf{FeynCalc/FormLink}~\cite{Mertig:1990an,Feng:2012tk}
to conduct the Dirac/color trace calculation.

The final state phase space implicit in (\ref{CS:def:Fragmentation:Function}) assumes the specific form~\cite{Bodwin:2003wh,Bodwin:2012xc}:
\begin{eqnarray}
d\Phi_n &=& {4\pi M_{h_c} \over S_n} \delta(k^+-P^+-\sum_{i=1}^n k_i^+) \prod_{i=1}^n \frac{dk^+_i}{2k_i^+}\frac{d^{D-2}k_{i\perp}}{(2\pi)^{D-1}} \theta(k^+_i)
\nn \\
&=& {4\pi M \over S_n} \delta(k^+-P^+-\sum_{i=1}^n k_i^+) \prod_{i=1}^n \frac{d^D k_i}{(2\pi)^{D}} 2\pi \delta_{+}(k^2_i), \label{phase:space}
\end{eqnarray}
where $k_i$ stands for the momentum of the $i$-th gluon in the final state, and 
$S_n$ is the statistical factor for $n$ identical gluons. For our purpose, suffices it
to know $S_1=1$ and $S_n=2$.

The ingenuity of \cite{Zhang:2017xoj} is to replace each $\delta$-function in (\ref{phase:space}) with the 
generalized cut propagator~\cite{Anastasiou:2002yz,Gehrmann-DeRidder:2003pne}, by invoking the
following identity:
\begin{equation}
\delta(x) = \frac{1}{2\pi i} \left[ \frac{1}{x-i\varepsilon} - \frac{1}{x+i\varepsilon} \right].
\label{cut:trick}
\end{equation}
One can then apply the integration-by-part (IBP) method to the phase-space integration just as in loop integration~\cite{Anastasiou:2002yz},
since the differentiation operation involved in IBP identities is insensitive to the $i\varepsilon$. 
Therefore, one can also utilize the packages \textsf{Apart}~\cite{Feng:2012iq} and \text{FIRE}~\cite{Smirnov:2014hma} to conduct partial fraction and the corresponding IBP reduction, and finally end up with a set of Master Integrals (MIs), which are much
easier to manipulate than the original integrand~\footnote{Note that in order to accomplish the overwhelmingly 
challenging calculation of the next-to-next-to-leading order QCD correction of
the $\eta_c$ hadronic width~\cite{Feng:2017hlu}, it is crucial to utilize this powerful trick.}.
Finally, we end up with nine MIs in the color-singlet channels and two MIs in the  color-octet channels.
The former class of MIs can be parameterized as
\beq
F(i,j,m,n) = \int d \Phi_2 \frac{1}{E_1^i} \frac{1}{E_2^j} \frac{1}{E_3^m} \frac{1}{E_4^n},
\eeq
where the propagators $E_i(i=1,\cdots,4)$ are
\beq
	E_{1,2} = (k_{1,2}+p)^2-m^2,\quad E_3 = (k_1+k_2+p)^2-m^2, \quad E_4 = (k_1+k_2+2p)^2,
\label{propagators:def}
\eeq
respectively. $p=P/2$ is the half of the $h_c$ momentum.
The 9 encountered MIs in the singlet channel are labeled by the indices:
\bqa
(i,j,m,n) &=& (0,0,1,0),\quad(0,0,0,1),\quad(1,1,0,0),\quad\,(0,1,2,0),\quad(0,1,1,0),
\nn \\
&=& (0,1,0,2),\quad (0,1,0,1),\quad (0,0,1,1),\quad (1,1,0,1).
\eqa
Most of them can be readily worked out by the standard method, that is,
rewriting the propagators in (\ref{propagators:def}) in terms of light-cone variables,
conducting two-loop integration in $D-2$-dimensional transverse momentum ${\bf k}_{1,2\perp}$,
then followed by a one-dimensional parametric integration.
The last MI, which involves three propagators, is however somewhat challenging. 
Fortunately, with the aid of the differential equation technique, 
its analytic expression has already been unravelled in \cite{Zhang:2017xoj}.

The two MIs in the color-octet channel are rather rudimentary,
\begin{equation}
\int d\Phi_1 \frac{1}{(k_1+p)^2} ,\quad \int d\Phi_1 \frac{1}{(k_1+p)^2-m^2}.
\end{equation}
 
NRQCD provides a systematic framework to enable one to factor the IR divergence encountered in the color-singlet
channel into the color-octet production matrix element $\langle {\cal O}_8^{h_c}(^1S_0)\rangle$, 
as originally demonstrated in $B$ meson decay to $P$-wave charmonium~\cite{Bodwin:1992qr}.
 
After some straightforward but tedious manipulation, and following the recipe given in Refs.~\cite{Beneke:1998ks,Jia:2012qx} to eliminate the IR singularity under the $\overline{\rm MS}$ factorization scheme,
we finally obtain the following SDCs: 
\begin{subequations}
\bqa
d_{1}(z,\mu_\Lambda)&=&\frac{\alpha_s^3 B_F}{3 \pi N_c^2}\Bigg\{
\Big[(3-2 z) z+2 (1-z) \ln(1-z)\Big] \ln\left(\frac{4 m^2}{\mu_\Lambda ^2}\right)
-\frac{14}{15} (6-5 z) \text{Li}_2\left(\frac{1-z}{2-z}\right)
\nn\\
&&+\frac{2}{15} \left(33-17 z-9 z^2\right) \text{Li}_2\left(\frac{2-2z}{2-z}\right)
+\frac{1}{15} (18-25 z) \text{Li}_2(1-z) -\frac{3+z}{8} I_9
\nn\\
&& -\frac{c_1}{60 (2-z)^2 (1-z) (1-2 z)^3} -\frac{c_2 \ln(1-z)}{60 (1-2 z)^4} +\frac{c_3 (1-z) \ln^2(1-z)}{120 (1-2 z)^5}
\\
&&-\frac{c_4 z \ln(2-z)}{30 (2-z)^2 (1-z)^2} -\frac{c_5 \ln^2(2-z)}{120 (1-z)^3} -\frac{c_6 z \ln z}{60 (2-z)^2 (1-z)^2 (1-2 z)^4}
\nn\\
&& +\frac{c_7 z \ln(1-z) \ln z}{60 (1-2 z)^5}  -\frac{c_8 z^2 \ln(2-z) \ln z}{60 (1-z)^3} -\frac{c_9 z \ln^2z}{120 (1-z)^3 (1-2 z)^5} \Bigg\},
\nn\\
d_8(z) &=& \frac{\alpha_s^2 B_F}{N_c  C_F} \Big[ (1-z)\ln(1-z) + \frac{1}{2}(3-2z)z \Big],
\eqa
\end{subequations}
where $B_F=(N_c^2-4)/(4N_c)$ and $c_i(i=1,\cdots,9)$ are defined as
\begin{subequations}
\bqa
c_1 &=& \pi^2 \left(112-960 z+3292 z^2-5684 z^3+4972 z^4-1628 z^5-504 z^6+496 z^7-96 z^8\right)
\nn\\
&& + 1584 z-14052 z^2+54785 z^3-119579 z^4+158947 z^5-131887 z^6
\\
&&+66880 z^7-18980 z^8+2304 z^9,
\nn\\
c_2 &=& 210-2093 z+7890 z^2-13501 z^3+5326 z^4+16632 z^5-26920 z^6
\\
&& +15408 z^7-3072 z^8,
\nn\\
c_3 &=& 408-4097 z+16233 z^2-30962 z^3+21898 z^4+20284 z^5-57056 z^6
\\
&& +54864 z^7-28256 z^8+6144 z^9,
\nn \\
c_4 &=& 38+920 z-4752 z^2+9402 z^3-9621 z^4+5417 z^5-1595 z^6+192 z^7,
\\
c_5 &=& 336-1288 z+1668 z^2+4 z^3-2770 z^4+4061 z^5-2991 z^6+1171 z^7-192 z^8,
\\
c_6 &=& 896-10108 z+54243 z^2-175908 z^3+380634 z^4-582264 z^5+651071 z^6
\\
&& -535798 z^7+316152 z^8-125336 z^9+29488 z^{10}-3072 z^{11},
\nn\\
c_7 &=& 45-210 z-645 z^2+10180 z^3-41842 z^4+89500 z^5-111920 z^6
\\
&& +83120 z^7-34400 z^8+6144 z^9,
\nn\\
c_8 &=& 180-1180 z+3050 z^2-4061 z^3+2991 z^4-1171 z^5+192 z^6,
\\
c_9 &=& 45-525 z+3100 z^2-10610 z^3+22054 z^4-28190 z^5+21555 z^6
\\
&& -8960 z^7+1500 z^8+32 z^9,
\nn
\eqa
\end{subequations}
and $I_9$ is defined as
\bqa
	I_9&=&\text{Li}_3\left(\frac{z}{z-1}\right)+\text{Li}_3\left(\frac{2
z-1}{z-1}\right)+\text{Li}_3\left(\frac{2z-1}{z}\right)
+\text{Li}_2\left(\frac{2
z-1}{z-1}\right) \ln\left(\frac{1-z}{z}\right)
\nn\\
&&-\text{Li}_2(z)\ln\left(\frac{1-z}{z}\right)  +\frac{1}{6}
\ln^3\left(\frac{1-z}{z}\right)
-\frac{1}{2} \ln (1-z)
\ln z \ln\left(\frac{1-z}{z}\right)-\zeta(3).
\eqa

From these analytic expressions, it is not difficult to
find the asymptotic behavior of
these short-distance coefficients near $z=1$:
\begin{subequations}
\bqa
  && d_1(z) \longrightarrow\frac{\alpha_s B_F}{3\pi N_c^2} \Bigg[ \ln\left(\frac{4m^2}{\mu_\Lambda^2}\right)+2\ln(1-z) +
  \frac{1}{3} \Bigg] + {\cal O}\left[(1-z)\ln^2(1-z)\right]\,
\\
  && d_8(z) \longrightarrow \frac{\alpha_s^2 B_F}{2 N_c C_F} + {\cal O}((1-z)\ln(1-z)).
\eqa
\end{subequations}
Notice the mild logarithmic singularity of $d_1(z)$ is developed in the $z\to 1$ limit.

It is also straightforward to obtain the fragmentation probability,
\begin{eqnarray}
 \int_0^1 D_{g \to h_c}(z,\mu) \, dz = \frac{\alpha_s^2 B_F}{6 N_c C_F} \frac{\langle {\cal O}_8^{h_c}(^1S_0)\rangle_{\mu_\Lambda}}{m^3} + \frac{\alpha_s^3 B_F}{3 \pi N_c^2} \frac{\langle {\cal O}_1^{h_c}(^1S_0)\rangle}{m^5} \left[ \frac{1}{3}\ln\left(\frac{4 m^2}{\mu_\Lambda^2}\right)-1.1341 \right].
 \nn\\
\label{frag:prob}
\end{eqnarray}

\begin{figure}[tb]
\centering
\includegraphics[width=0.6\textwidth]{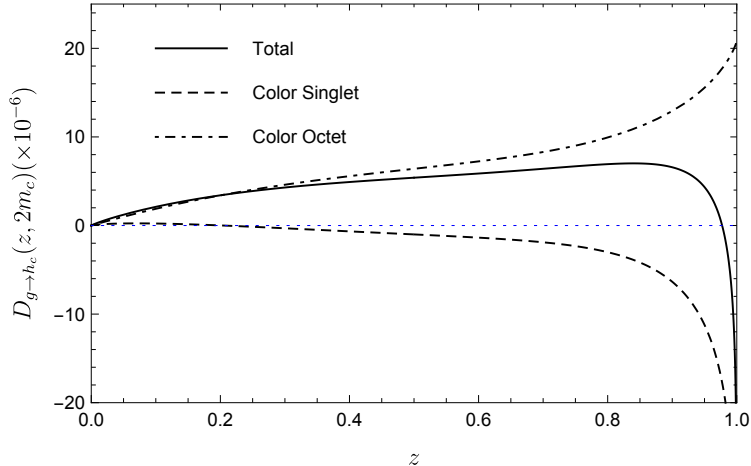}
\caption{The gluon fragmentation function $D_{g\to h_c}(z)$ evaluated at $\mu_\Lambda=m_c$.
\label{figDg}}
\end{figure}

For concreteness, we take the following input parameters:
\bqa
&&	m_c=1.5\,\text{GeV},\;\langle {\cal O}_1^{h_c}(^1P_1) = 0.322\,\text{GeV}^5, \; \langle {\cal O}_8^{h_c}(^1S_0)(m) \rangle = 0.02\,\text{GeV}^3, \nn\\
&& \mu_\Lambda=m_c,\; \alpha_s(2m_c)=0.26.
\eqa
Substituting these values into (\ref{frag:prob}), we then find the total fragmentation probability is 
about $(4.50\times10^{-6})$. The profile of the gluon fragmentation function is displayed in Fig.~\ref{figDg}.

Inspired by the recent technical advance in computating the quarkonium fragmentation function~\cite{Zhang:2017xoj},
in this work, we revisit the gluon-to-$h_c$ fragmentation function, and, for the first time
achieve the analytical, gauge-invariant expression for this FF. Our study might shed some light on the future establishing of
the $h_c$ and $h_b$ states at the \textsf{LHC} experiment.

\begin{acknowledgments}
The work of F.~F. is supported by the National Natural Science Foundation of China under Grant No.~11505285,
and by the Fundamental Research Funds for the Central Universities.
The work of S.~I, Y.~J. and J.~Z.
is supported in part by the National Natural Science Foundation of China under Grants No.~11475188,
No.~11621131001 (CRC110 by DGF and NSFC), by the IHEP Innovation Grant under contract number Y4545170Y2,
and by the State Key Lab for Electronics and Particle Detectors.
S.~I. also wishes to acknowledge the financial support from the
CAS-TWAS President's Fellowship Programme.
The Feynman diagrams in this paper are prepared by using \textsf{JaxoDraw}~\cite{Binosi:2003yf}.
\end{acknowledgments}

\end{document}